\documentclass[aps,twocolumn,showpacs]{revtex4}
\usepackage{amsmath}
\usepackage{epsfig}
\usepackage{multirow}

\begin{document}
	
\title{Molecular state interpretation of charmed baryons in the quark model}
\author{Ye Yan$^1$}\email{221001005@njnu.edu.cn}
\author{Xiaohuang Hu$^1$}\email{191002007@njnu.edu.cn}
\author{Yuheng Wu$^1$}\email{201001002@njnu.edu.cn}
\author{Hongxia Huang$^1$}\email{hxhuang@njnu.edu.cn(Corresponding author)}
\author{Jialun Ping$^1$}\email{jlping@njnu.edu.cn(Corresponding author)}
\author{Youchang Yang$^2$}\email{yangyc@gues.edu.cn}
\affiliation{$^1$Department of Physics, Nanjing Normal University, Nanjing, Jiangsu 210097, P. R. China}
\affiliation{$^2$School of Physics and Electronic Science, Zunyi Normal University, Zunyi 563006, P.R. China}

\begin{abstract}
Stimulated by the observation of $\Lambda_c(2910)^+$ by the Belle Collaboration, the $S$-wave $qqq\bar{q}c~(q=u~\text{or}~d)$ pentaquark systems with $I$ = 0, $J^P$ = $\frac{1}{2}^-,~\frac{3}{2}^- and~\frac{5}{2}^-$ are investigated in the framework of quark delocalization color screening model(QDCSM).
The real-scaling method is utilized to check the bound states and the genuine resonance states. The root mean square of cluster spacing is also calculated to study the structure of the states and estimate if the state is resonance state or not. The numerical results show that $\Lambda_{c}(2910)$ cannot be interpreted as a molecular state, and $\Sigma_{c}(2800)$ cannot be explained as the $ND$ molecular state with $J^P=1/2^-$. $\Lambda_{c}(2595)$ can be interpreted as the molecular state with $J^P=\frac{1}{2}^-$ and the main component is $\Sigma_{c}\pi$. $\Lambda_{c}(2625)$ can be interpreted as the molecular state with $J^P=\frac{3}{2}^-$ and the main component is $\Sigma_{c}^{*}\pi$. $\Lambda_{c}(2940)$ is likely to be interpreted as a molecular state with $J^P=3/2^-$, and the main component is $ND^{*}$. Besides, two new molecular states are predicted, one is the $J^P=3/2^-$ $\Sigma_{c}\rho$ resonance state with the mass around 3140 MeV, another one is the $J^P=\frac{5}{2}^-$ $\Sigma_{c}^*\rho$ with the mass of 3188.3 MeV.
\end{abstract}
	
	
\maketitle

\setcounter{totalnumber}{5}
	
\section{\label{sec:introduction}Introduction}
Not long ago, the Belle Collaboration reported that a new structure was found in the  $M_{\Sigma_{c}(2455)^{0,++}\pi^{\pm}}$  spectrum with a significance of  $4.2 \sigma$  including systematic uncertainty, which is tentatively named  $\Lambda_{c}(2910)^{+}$.
Its mass and width are measured to be $(2913.8  \pm   5.6 \pm 3.8) \mathrm{MeV} / c^{2}$  and  $(51.8 \pm 20.0 \pm 18.8) \mathrm{MeV}$, respectively~\cite{Belle:2022hnm}.
So far, there have been some theoretical analyses of this state.
In Ref.~\cite{Azizi:2022dpn}, the author utilized the light-cone QCD sum rule and interpreted that this baryon is a 2$P$ state with $J^P=1/2^-$ denoting by $\Lambda_{c}(1/2^-,2P)$.
In Ref.~\cite{Wang:2022dmw}, the author concluded that the newly observed $\Lambda_{c}(2910)$ can be explained as the $J^P=5/2^-$ state $\Lambda_{c}\left|J^{P}=\frac{5}{2}^{-}, 2\right\rangle_{\rho}$ in the framework of the chiral quark model.
While in Ref.~\cite{Zhang:2022pxc}, they used an unquenched picture to study $\Lambda_{c}(2910)$ by considering $S-$wave $D^*N$ channel coupled with the bare $udc$ core $(\Lambda_{c}(2P))$.
From their results, $\Lambda_{c}(2910)$ is deemed to contain a significant $D^*N$ component, and the bare state can cause the $D^*N$ binding more compactly.

Since $\Lambda_{c}^+$~\cite{Knapp:1976qw} was first observed by Fermilab in 1976, charmed baryon family has been enriched step by step with the help of experimental collaborations~\cite{ARGUS:1993vtm,CLEO:1994oxm,ARGUS:1997snv,E687:1993bax,CLEO:2000mbh,LHCb:2017jym,BaBar:2006itc,Belle:2021qip,Belle:2014fde,Ammosov:1993pi,CLEO:1996czm,Belle:2004zjl,LHCb:2020gge,ALICE:2021bli,Belle:2016lhy,CLEO:1998wvk,Belle:2013htj,Belle:2020ozq,LHCb:2020iby,Belle:2020tom,Belle:2016tai,Belle:2021gtf,BaBar:2006pve,LHCb:2021ptx,LHCb:2017uwr,Belle:2017ext}.
The experimental observations of charmed baryons have stimulated broad interest in understanding the structures of these states among theoretical groups~\cite{Hofmann:2005sw,He:2006is,Chen:2007xf,Garcilazo:2007eh,Valcarce:2008dr,Ebert:2007nw,Zhong:2007gp,Garcia-Recio:2008rjt,He:2010zq,Dong:2010gu,Dong:2010xv,Dong:2009tg,Ebert:2011kk,Wang:2011zzw,Haidenbauer:2010ch,Romanets:2012hm,Ortega:2012cx,Zhang:2014ska,Zhang:2012jk,Yasui:2014cwa,Dong:2014ksa,Chen:2014nyo,Cheng:2015naa,Chen:2015kpa,Lu:2014ina,Yoshida:2015tia,Wang:2015rda,Entem:2016lzh,Lu:2016gev,Zhao:2016zhf,Shah:2016mig,Guo:2016wpy,Arifi:2017sac,Chen:2016iyi,Nagahiro:2016nsx,Chen:2017sci,Wang:2017kfr,Dong:2017gaw,Lu:2018utx,Liang:2014kra,Yao:2018jmc,Guo:2019ytq,Nieves:2019nol,Lu:2019rtg,Huang:2016ygf,Wang:2020dhf,Niu:2020gjw,Sakai:2020psu,Luo:2019qkm,Kim:2020imk,Arifi:2021orx,Yang:2021lce,Gong:2021jkb,Zhang:2020dwp,Yu:2022ymb,Garcia-Tecocoatzi:2022zrf,Wang:2021bmz,Niu:2021qcc}, especially these excited charmed baryons such as $\Lambda_{c}(2595)$, $\Lambda_{c}(2625)$, $\Sigma_{c}(2800)$ and $\Lambda_{c}(2940)$. Considering that these charmed baryons contain both light and heavy quarks, studying these systems could provide a transitional link between light and heavy baryons.
At the same time, analysing the properties of these states could also deepen our understanding of the non-perturbative behavior of quantum chromodynamics (QCD).

In the last two decades, the focus of the various theoretical work and the controversial point is whether these states are excitations or multi-quark states.
We can briefly review some of the theoretical work that concerning the single charmed baryons mentioned earlier.
These states are studied by assuming that they are traditional three-quark excitations in the constituent quark model~\cite{Garcilazo:2007eh,Chen:2016iyi,Yoshida:2015tia},
the relativistic flux tube model~\cite{Wang:2011zzw,Chen:2014nyo},
the non-relativistic quark model~\cite{Nagahiro:2016nsx},
the relativistic quark-diquark model~\cite{Cheng:2015naa,Ebert:2007nw,Ebert:2011kk,Shah:2016mig,Yu:2022ymb},
the chiral quark model~\cite{Zhong:2007gp,Wang:2021bmz},
the QCD sum rules~\cite{Chen:2015kpa,Yang:2021lce,Chen:2017sci},
the effective field theory~\cite{Nieves:2019nol,Wang:2020dhf},
the effective Lagrangian method~\cite{Arifi:2017sac},
the Faddeev method in momentum space~\cite{Valcarce:2008dr}, and so on.

On the other hand, the multi-quark interpretations of these states were also investigated in the framework of an unitary baryon-meson coupled-channel model~\cite{Garcia-Recio:2008rjt,Romanets:2012hm},
the constituent quark model~\cite{Ortega:2012cx,Entem:2016lzh,Zhang:2020dwp,Zhao:2016zhf},
the one boston exchange model~\cite{He:2006is,He:2010zq},
the QCD sum rules~\cite{Zhang:2012jk,Zhang:2014ska},
the effective field theory~\cite{Nieves:2019nol,Sakai:2020psu},
a general framework that goes beyond effective range expansion~\cite{Guo:2016wpy},
the effective Lagrangian approach~\cite{Wang:2015rda},
the meson-exchange picture~\cite{Haidenbauer:2010ch},
the unitarized chiral perturbation theory~\cite{Lu:2014ina,Lu:2016gev},
the heavy quark spin symmetry~\cite{Liang:2014kra}, and so on.

In addition to the studies of masses and structures, the studies of the decay of these states were carried out in the effective meson Lagrangian~\cite{Dong:2010gu,Dong:2010xv,Dong:2009tg},
the $^3P_0$ strong decay model~\cite{Lu:2018utx,Guo:2019ytq,Gong:2021jkb,Garcia-Tecocoatzi:2022zrf},
and the constituent quark model~\cite{Wang:2017kfr}. Moreover, the view of these states as a mixture of three-quark and five-quark were considered in an unquenched picture~\cite{Luo:2019qkm}.

For $\Lambda_{c}(2595)$, it was investigated as multi-quark state in Ref.~\cite{Nieves:2019nol}, where $\Lambda_{c}(2595)$ was predicted to have a predominant molecular structure with the help the effective field theory. This is because it is either the result of the chiral $\Sigma_{c}\pi$ interaction, whose threshold is located much closer than the mass of the bare three-quark state, or because the light degrees of freedom in its inner structure are coupled to the unnatural $0^-$ quantum numbers.
Meanwhile $\Lambda_{c}(2595)$ was also studied in terms of traditional three-quark state.
For the low-lying $\Lambda_{c}(2595)$ baryon, the non-relativistic quark model description as the $\lambda$-mode excitation with a spin-0 diquark can explain the decay property well in Ref.~\cite{Nagahiro:2016nsx}.

As for $\Lambda_{c}(2625)$, in Ref.~\cite{Liang:2014kra}, a state with spin 3/2 that couples mostly to $D^{*}N$ was associated to the experimental found $\Lambda_{c}(2625)$ in the framework of the heavy quark spin symmetry.
Meanwhile, $\Lambda_{c}(2625)$ can be explained as an excited three-quark state in Ref.~\cite{Garcia-Tecocoatzi:2022zrf}, where $\Lambda_{c}(2625)$ was identified as a $P_{\lambda}$-wave excitation, with $J^P=3/2^-$ and $S=1/2$.

According to Ref.~\cite{Zhang:2012jk}, the author investigated $\Sigma_{c}(2800)$ as the $S$-wave $DN$ state with $J^P=1/2^-$ in the framework of QCD sum rules.
However, the research that regards $\Sigma_{c}(2800)$ as a traditional three-quark state was carried out in Ref.~\cite{Wang:2020dhf}, where the explanation of $\Sigma_{c}(2800)$ as the $DN$ molecular state was disfavored. In Ref.~\cite{Wang:2020dhf}, $\Sigma_{c}(2800)$ was more likely to be the conventional $1P$ charmed baryon, since its mass was well consistent with the quark model prediction.

$\Lambda_{c}(2940)$ was regarded as multi-quark state and studied in Ref.~\cite{Ortega:2012cx}.
They proposed a theoretical explanation of the $\Lambda_{c}(2940)$ as a molecular state in a constituent quark model that has been extensively used to describe hadron phenomenology.
However, the work of Ref.~\cite{Valcarce:2008dr} used the Faddeev method in momentum space and considered $\Lambda_{c}(2940)$ being the first radial excitation $2S$ of the $\Sigma_{c}$ with $J^P=3/2^+$.

An alternative approach to study hadron-hadron interaction and the multi-quark states is the quark delocalization color screening model(QDCSM), which was developed in the 1990s with the aim of explaining the similarities between nuclear and molecular forces~\cite{Wu:1996fm}. The model gives a good description of $NN$ and $YN$ interactions and the properties of deuteron~\cite{Ping:2000dx,Ping:1998si,Wu:1998wu,Pang:2001xx}. It is also employed to calculate the baryon-baryon and baryon-meson scattering phase shifts in the framework of the resonating group method (RGM), and the exotic hadronic states are also studied in this model. Studies also show that the $NN$ intermediate-range attraction mechanism in the QDCSM, quark delocalization, and color screening, is equivalent to the $\sigma$-meson exchange in the chiral quark model, and the color screening is an effective description of the hidden-color channel coupling~\cite{ChenLZ,Huang:2011kf}. So it is feasible and meaningful to extend this model to investigate the charmed baryons.

In this work, we explore the molecular state interpretation of charmed baryons in QDCSM. Both $qqq-\bar{q}c$ and $qqc-\bar{q}q$ structures, as well as the coupling of these two structures are taken into account.
Our purpose is to investigate whether $\Lambda_c(2910)$ could be explained as a molecular state. In addition, we also want to see if any other bound or resonance state exist or not.
Besides, with the help of real-scaling method, we can confirm these possible bound states and the genuine resonance states.

This paper is organized as follows. After introduction, we briefly introduce the quark model and methods in section II.
Then, the numerical results and discussions are presented in Section III.
Finally, the paper ends with summary in Section IV.

\section{THEORETICAL FRAMEWORK}
Herein, QDCSM is employed to investigate the properties of $qqq\bar{q}c$ systems, and the channel coupling effect is considered.
In this sector, we will introduce this model and the way of constructing wave functions.
	
\subsection{Quark delocalization color screening model (QDCSM)}

The QDCSM is an extension of the native quark cluster model~\cite{DeRujula:1975qlm,Isgur:1978xj,Isgur:1978wd,Isgur:1979be}.
It has been developed to address multi-quark systems.
The detail of QDCSM can be found in Refs.~\cite{Wu:1996fm,Huang:2011kf,Ping:1998si,Wu:1998wu,Pang:2001xx,Ping:2000cb,Ping:2000dx,Ping:2008tp}.
Here, we mainly present the salient features of the model.
The general form of the pentaquark Hamiltonian is given by
	\begin{align}
		H=&\sum_{i=1}^5\left(m_i+\frac{\boldsymbol{p}_{i}^{2}}{2m_i}\right)-T_{CM} +\sum_{j>i=1}^5 V(\boldsymbol{r}_{ij})
	\end{align}
where $m_i$ is the quark mass, $\boldsymbol{p}_{i}$ is the momentum of the quark, and $T_{CM}$ is the center-of-mass kinetic energy.
The dynamics of the pentaquark system is driven by a two-body potential
	\begin{align}
		V(\boldsymbol{r}_{ij})= & V_{CON}(\boldsymbol{r}_{ij})+V_{OGE}(\boldsymbol{r}_{ij})+V_{\chi}(\boldsymbol{r}_{ij})
	\end{align}
The most relevant features of QCD at its low energy regime: color confinement ($V_{CON}$), perturbative one-gluon exchange interaction ($V_{OGE}$), and dynamical chiral symmetry breaking ($V_{\chi}$) have been taken into consideration.

Here, a phenomenological color screening confinement potential($V_{CON}$) is used as
\begin{align}
	V_{CON}(\boldsymbol{r}_{ij}) = & -a_{c}\boldsymbol{\lambda}_{i}^{c} \cdot \boldsymbol{\lambda}_{j}^{c}\left[  f(\boldsymbol{r}_{ij})+V_{0}\right] ,
\end{align}
\begin{align}
	f(\boldsymbol{r}_{ij}) =& \left\{\begin{array}{l}
		\boldsymbol{r}_{i j}^{2} ~~~~~~~~~~~~~ ~i,j ~\text {occur in the same cluster } \\
		\frac{1-e^{-\mu_{q_{i}q_{j}} \boldsymbol{r}_{i j}^{2}}}{\mu_{q_{i}q_{j}}}  ~~~i,j ~\text {occur in different cluster }
	\end{array}\right.
\end{align}
where $a_c$, $V_{0}$ and $\mu_{q_{i}q_{j}}$ are model parameters, and $\boldsymbol{\lambda}^{c}$ stands for the SU(3) color Gell-Mann matrices.
Among them, the color screening paprameter $\mu_{q_{i}q_{j}}$ is determined by fitting the deuteron properties, nucleon-nucleon scattering phase shifts, and hyperon-nucleon scattering phase shifts, respectively, with $\mu_{qq}=0.45~$fm$^{-2}$, $\mu_{qs}=0.19~$fm$^{-2}$ and $\mu_{ss}=0.08~$fm$^{-2}$, satisfying the relation, $\mu_{qs}^{2}=\mu_{qq}\mu_{ss}$~\cite{ChenM}.
Besides, we found that the heavier the quark, the smaller this parameter $\mu_{q_{i}q_{j}}$.
When extending to the heavy quark system, the hidden-charm pentaqyark system, we took $\mu_{cc}$ as a adjustable parameter from $0.01~$fm$^{-2}$ to $0.001~$fm$^{-2}$, and found that the results were insensitive to the value of $\mu_{cc}$~\cite{HuangPc1}.
Moreover, the $P_{c}$ states were well predicted in the work of Refs.~\cite{HuangPc1,HuangPc2}.
So here we take $\mu_{cc}=0.01~$fm$^{-2}$ and $\mu_{qc}=0.067~$fm$^{-2}$, also satisfy the relation $\mu_{qc}^{2}=\mu_{qq}\mu_{qc}$.

In the present work, we mainly focus on the low-lying negative parity $qqq\bar{q}c$ pentaquark states of $S$-wave, so the spin-orbit and tensor interactions are not included.
The one-gluon exchange potential ($V_{OGE}$), which includes coulomb and color-magnetic interactions, is written as
	\begin{align}
		V_{OGE}(\boldsymbol{r}_{ij})= &\frac{1}{4}\alpha_{s} \boldsymbol{\lambda}_{i}^{c} \cdot \boldsymbol{\lambda}_{j}^{c}  \\
		&\cdot \left[\frac{1}{r_{i j}}-\frac{\pi}{2} \delta\left(\mathbf{r}_{i j}\right)\left(\frac{1}{m_{i}^{2}}+\frac{1}{m_{j}^{2}}+\frac{4 \boldsymbol{\sigma}_{i} \cdot \boldsymbol{\sigma}_{j}}{3 m_{i} m_{j}}\right)\right]   \nonumber \label{Voge}
	\end{align}
where $\boldsymbol{\sigma}$ is the Pauli matrices and $\alpha_{s}$ is the quark-gluon coupling constant.

However, the quark-gluon coupling constant between quark and anti-quark, which offers a consistent description of mesons from light to heavy-quark sector, is determined by the mass differences between pseudoscalar mesons (spin-parity $J^P=0^-$) and vector (spin-parity $J^P=1^-$), respectively.
For example, from the model Hamiltonian, the mass difference between $D$ and $D^*$ is determined by the color-magnetic interaction in Eq.(\ref{Voge}), so the parameter $\alpha_s(qc)$ is determined by fitting the mass difference between $D$ and $D^*$.

The dynamical breaking of chiral symmetry results in the SU(3) Goldstone boson exchange interactions appear between constituent light quarks $u, d$ and $s$.
Hence, the chiral interaction is expressed as
\begin{align}
	V_{\chi}(\boldsymbol{r}_{ij})= & V_{\pi}(\boldsymbol{r}_{ij})+V_{K}(\boldsymbol{r}_{ij})+V_{\eta}(\boldsymbol{r}_{ij})
\end{align}
Among them
\begin{align}
V_{\pi}\left(\boldsymbol{r}_{i j}\right) =&\frac{g_{c h}^{2}}{4 \pi} \frac{m_{\pi}^{2}}{12 m_{i} m_{j}} \frac{\Lambda_{\pi}^{2}}{\Lambda_{\pi}^{2}-m_{\pi}^{2}} m_{\pi}\left[Y\left(m_{\pi} \boldsymbol{r}_{i j}\right)\right. \nonumber \\
&\left.-\frac{\Lambda_{\pi}^{3}}{m_{\pi}^{3}} Y\left(\Lambda_{\pi} \boldsymbol{r}_{i j}\right)\right]\left(\boldsymbol{\sigma}_{i} \cdot \boldsymbol{\sigma}_{j}\right) \sum_{a=1}^{3}\left(\boldsymbol{\lambda}_{i}^{a} \cdot \boldsymbol{\lambda}_{j}^{a}\right)
\end{align}
\begin{align}
	V_{K}\left(\boldsymbol{r}_{i j}\right) =&\frac{g_{c h}^{2}}{4 \pi} \frac{m_{K}^{2}}{12 m_{i} m_{j}} \frac{\Lambda_{K}^{2}}{\Lambda_{K}^{2}-m_{K}^{2}} m_{K}\left[Y\left(m_{K} \boldsymbol{r}_{i j}\right)\right. \nonumber \\
	&\left.-\frac{\Lambda_{K}^{3}}{m_{K}^{3}} Y\left(\Lambda_{K} \boldsymbol{r}_{i j}\right)\right]\left(\boldsymbol{\sigma}_{i} \cdot \boldsymbol{\sigma}_{j}\right) \sum_{a=4}^{7}\left(\boldsymbol{\lambda}_{i}^{a} \cdot \boldsymbol{\lambda}_{j}^{a}\right)
\end{align}
\begin{align}
	V_{\eta}\left(\boldsymbol{r}_{i j}\right) =&\frac{g_{c h}^{2}}{4 \pi} \frac{m_{\eta}^{2}}{12 m_{i} m_{j}} \frac{\Lambda_{\eta}^{2}}{\Lambda_{\eta}^{2}-m_{\eta}^{2}} m_{\eta}\left[Y\left(m_{\eta} \boldsymbol{r}_{i j}\right)\right. \nonumber \\
	&\left.-\frac{\Lambda_{\eta}^{3}}{m_{\eta}^{3}} Y\left(\Lambda_{\eta} \boldsymbol{r}_{i j}\right)\right]\left(\boldsymbol{\sigma}_{i} \cdot \boldsymbol{\sigma}_{j}\right)\left[\cos \theta_{p}\left(\boldsymbol{\lambda}_{i}^{8} \cdot \boldsymbol{\lambda}_{j}^{8}\right)\right.  \nonumber \\
	&\left.-\sin \theta_{p}\left(\boldsymbol{\lambda}_{i}^{0} \cdot \boldsymbol{\lambda}_{j}^{0}\right)\right]
\end{align}
where $Y(x) = e^{-x}/x$ is the standard Yukawa function.
The physical $\eta$ meson is considered by introducing the angle $\theta_{p}$ instead of the octet one. The $\boldsymbol{\lambda}^a$ are the SU(3) flavor Gell-Mann matrices.
The values of $m_\pi$, $m_k$ and $m_\eta$ are the masses of the SU(3) Goldstone bosons, which adopt the experimental values~\cite{ParticleDataGroup:2020ssz}.
The chair coupling constant $g_{ch}$, is determined from the $\pi N N$ coupling constant through
\begin{align}
	\frac{g_{c h}^{2}}{4 \pi} & = \left(\frac{3}{5}\right)^{2} \frac{g_{\pi N N}^{2}}{4 \pi} \frac{m_{u, d}^{2}}{m_{N}^{2}}
\end{align}

Asssuming that flavor SU(3) is an exact symmetry, it will only be broken by the different mass of the strange quark.
As we can see, in the present work, there exits no strange quark in the $qq\bar{q}c$ systems.
As a result, the K-meson exchange potential will have no effect on the energies of the possible states.

The other symbols in the above expressions have their usual meanings.
All the parameters shown in Table~\ref{parameters} are fixed by masses of the ground baryons and mesons. Table~\ref{hadrons} shows the masses of the baryons and mesons used in this work.
\begin{table}[ht]
	\caption{\label{parameters}Model parameters:
		$m_{\pi}=0.7$ fm$^{-1}$,
		$m_{K}=2.51$ fm$^{-1}$,
		$m_{\eta}=2.77$ fm$^{-1}$,
		$\Lambda_{\pi}=4.2$ fm$^{-1}$,
		$\Lambda_{K}=5.2$ fm$^{-1}$,
		$\Lambda_{\eta}=5.2$ fm$^{-1}$,
		$\frac{g_{ch}^2}{4\pi}$=0.54.}
	\begin{tabular}{cccccc} \hline\hline
		~~~~$b$~~~~ & ~~$m_{u,d}$~~ & ~~~$m_{c}$~~~  & ~~~$V_{0_{qq}}$~~~~&~~~$V_{0_{\bar{q}q}}$~~~~& ~~~$ a_c$~~~   \\
		(fm) & (MeV) & (MeV)  & (fm$^{-2}$)  & (fm$^{-2}$) & ~(MeV\,fm$^{-2}$)~  \\
		0.518  & 313 & 1788   &    -1.288 &  -0.743  &  58.03 \\ \hline
		$\alpha_{s_{uu}}$ &  $\alpha_{s_{uc}}$ & $\alpha_{s_{u\bar{u}}}$ & $\alpha_{s_{u\bar{c}/c\bar{u}}}$ & &\\
		0.565 & 0.467  & 1.491 & 1.200 & &   \\ \hline\hline
	\end{tabular}
\end{table}
\begin{table}[ht]
	\caption{The masses (in MeV) of the baryons and mesons used in this work. Experimental values are taken
		from the Particle Data Group (PDG)~\cite{ParticleDataGroup:2020ssz}.}
	\begin{tabular}{cccc}
		\hline \hline
		~~~~Hadron~~~~& ~~~~~$I(J^P)$~~~~~  & ~~~~~~$M_{the.}$~~~~~ & ~~~~~$M_{exp.}$~~~~ \\ \hline
		$N$          & $1/2(1/2^+)$ & 939   & 939 \\
		$\Delta$     & $3/2(3/2^+)$ & 1232  & 1232 \\
		$\Sigma_c$   & $1(1/2^+)$   & 2465  & 2455 \\
		$\Sigma^*_c$ & $1(3/2^+)$   & 2490  & 2518 \\
		$\Lambda_c$  & $0(1/2^+)$   & 2286  & 2286 \\
		$\pi$        & $1(0^-)$     & 139   & 139 \\
		$\rho$       & $1(1^-)$     & 770   & 770 \\
		$\omega$     & $0(1^-)$     & 722   & 782 \\
		$D$          & $1/2(0^-)$   & 1868  & 1869 \\
		$D^*$        & $1/2(1^-)$   & 1952  & 2007 \\  \hline\hline
	\end{tabular}
	\label{hadrons}
\end{table}
\subsection{Resonating group method and wave functions}
The resonating group method (RGM)~\cite{RGM1,RGM} and generating coordinates method~\cite{GCM1,GCM2} are used to carry out a dynamical calculation.
The main feature of the RGM for two-cluster systems is that it assumes that two clusters are frozen inside, and only considers the relative motion between the two clusters.
So the conventional ansatz for the two-cluster wave functions is
\begin{equation}
	\psi_{5q} = {\cal A }\left[[\phi_{B}\phi_{M}]^{[\sigma]IS}\otimes\chi(\boldsymbol{R})\right]^{J} \label{5q}
\end{equation}
where the symbol ${\cal A }$ is the anti-symmetrization operator, and ${\cal A } = 1-P_{14}-P_{24}-P_{34}$. $[\sigma]=[222]$ gives the total color symmetry and all other symbols have their usual meanings.
 $\phi_{B}$ and $\phi_{M}$ are the $q^{3}$ and $\bar{q}q$ cluster wave functions, respectively.
 From the variational principle, after variation with respect to the relative motion wave function $\chi(\boldsymbol{\mathbf{R}})=\sum_{L}\chi_{L}(\boldsymbol{\mathbf{R}})$, one obtains the RGM equation:
\begin{equation}
	\int H(\boldsymbol{\mathbf{R}},\boldsymbol{\mathbf{R'}})\chi(\boldsymbol{\mathbf{R'}})d\boldsymbol{\mathbf{R'}}=E\int N(\boldsymbol{\mathbf{R}},\boldsymbol{\mathbf{R'}})\chi(\boldsymbol{\mathbf{R'}})d\boldsymbol{\mathbf{R'}}  \label{RGM eq}
\end{equation}
where $H(\boldsymbol{\mathbf{R}},\boldsymbol{\mathbf{R'}})$ and $N(\boldsymbol{\mathbf{R}},\boldsymbol{\mathbf{R'}})$ are Hamiltonian and norm kernels.
By solving the RGM equation, we can get the energies $E$ and the wave functions.
In fact, it is not convenient to work with the RGM expressions.
Then, we expand the relative motion wave function $\chi(\boldsymbol{\mathbf{R}})$ by using a set of gaussians with different centers
\begin{align}
	\chi(\boldsymbol{R}) =& \frac{1}{\sqrt{4 \pi}}\left(\frac{6}{5 \pi b^{2}}\right)^{3 / 4} \sum_{i,L,M} C_{i,L} \\ \nonumber    
	&\cdot\int \exp \left[-\frac{3}{5 b^{2}}\left(\boldsymbol{R}-\boldsymbol{S}_{i}\right)^{2}\right] Y_{LM}\left(\hat{\boldsymbol{S}}_{i}\right) d \Omega_{\boldsymbol{S}_{i}}
\end{align}
where $L$ is the orbital angular momentum between two clusters, and $\boldsymbol {S_{i}}$, $i=1,2,...,n$ are the generator coordinates, which are introduced to expand the relative motion wave function. By including the center of mass motion:
\begin{equation}
	\phi_{C} (\boldsymbol{R}_{C}) = (\frac{5}{\pi b^{2}})^{3/4}e^{-\frac{5\boldsymbol{R}^{2}_{C}}{2b^{2}}}
\end{equation}
the ansatz Eq.(\ref{5q}) can be rewritten as
\begin{align}
	\psi_{5 q} =& \mathcal{A} \sum_{i,L} C_{i,L} \int \frac{d \Omega_{\boldsymbol{S}_{i}}}{\sqrt{4 \pi}} \prod_{\alpha=1}^{3} \phi_{\alpha}\left(\boldsymbol{S}_{i}\right) \prod_{\beta=4}^{5} \phi_{\beta}\left(-\boldsymbol{S}_{i}\right) \nonumber \\
	& \cdot\left[\left[\chi_{I_{1} S_{1}}\left(B\right) \chi_{I_{2} S_{2}}\left(M\right)\right]^{I S} Y_{LM}\left(\hat{\boldsymbol{S}}_{i}\right)\right]^{J} \nonumber \\
	& \cdot\left[\chi_{c}\left(B\right) \chi_{c}\left(M\right)\right]^{[\sigma]} \label{5q2}
\end{align}
where $\chi_{I_{1}S_{1}}$ and $\chi_{I_{2}S_{2}}$ are the product of the flavor and spin wave functions, and $\chi_{c}$ is the color wave function. These will be shown in detail later.  $\phi_{\alpha}(\boldsymbol{S}_{i})$ and $\phi_{\beta}(-\boldsymbol{S}_{i})$ are the single-particle orbital wave functions with different reference centers:
\begin{align}
	\phi_{\alpha}\left(\boldsymbol{S}_{i}\right) & = \left(\frac{1}{\pi b^{2}}\right)^{3 / 4} e^{-\frac{1}{2 b^{2}}\left(r_{\alpha}-\frac{2}{5} \boldsymbol{S}_{i}\right)^{2}}, \\ \nonumber
	\phi_{\beta}\left(\boldsymbol{-S}_{i}\right) & = \left(\frac{1}{\pi b^{2}}\right)^{3 / 4} e^{-\frac{1}{2 b^{2}}\left(r_{\beta}+\frac{3}{5} \boldsymbol{S}_{i}\right)^{2}}
\end{align}
With the reformulated ansatz Eq.(\ref{5q2}), the RGM Eq.(\ref{RGM eq}) becomes an algebraic eigenvalue equation:
\begin{equation}
	\sum_{j} C_{j}H_{i,j}= E \sum_{j} C_{j}N_{i,j}
\end{equation}
where $H_{i,j}$ and $N_{i,j}$ are the Hamiltonian matrix elements and overlaps, respectively.
By solving the generalized eigen problem, we can obtain the energy and the corresponding wave functions of the pentaquark systems.

The quark delocalization in QDCSM is realized by specifying the single particle orbital wave function of QDCSM as a linear combination of left and right Gaussians, the single particle orbital wave functions used in the ordinary quark cluster model
\begin{eqnarray}
	\psi_{\alpha}(\boldsymbol {S_{i}} ,\epsilon) & = & \left(
	\phi_{\alpha}(\boldsymbol {S_{i}})
	+ \epsilon \phi_{\alpha}(-\boldsymbol {S_{i}})\right) /N(\epsilon), \nonumber \\
	\psi_{\beta}(-\boldsymbol {S_{i}} ,\epsilon) & = &
	\left(\phi_{\beta}(-\boldsymbol {S_{i}})
	+ \epsilon \phi_{\beta}(\boldsymbol {S_{i}})\right) /N(\epsilon), \nonumber \\
	N(S_{i},\epsilon) & = & \sqrt{1+\epsilon^2+2\epsilon e^{-S_i^2/4b^2}} \label{1q}
\end{eqnarray}
Here, the mixing parameter $\epsilon$ is not an adjusted one but determined variationally by the dynamics of the multi-quark system itself.
In this way, the multi-quark system chooses its favorable configuration in the interacting process.
This mechanism has been used to explain the cross-over transition between hadron phase and quark-gluon plasma phase~\cite{Xu}.

For the spin wave function, we first construct the spin wave functions of the $q^{3}$ and $\bar{q}q$ clusters with SU(2) algebra, and then the total spin wave function of the pentaquark system is obtained by coupling the spin wave functions of two clusters together.
The spin wave functions of the $q^{3}$ and $\bar{q}q$ clusters are Eq.(19) and Eq.(20), respectively
	\begin{align}
		\chi_{\frac{3}{2}, \frac{3}{2}}^{\sigma}(3) & = \alpha \alpha \alpha \nonumber \\
		\chi_{\frac{3}{2}, \frac{1}{2}}^{\sigma}(3) & = \frac{1}{\sqrt{3}}(\alpha \alpha \beta+\alpha \beta \alpha+\beta \alpha \alpha)   \nonumber \\
		\chi_{\frac{3}{2},-\frac{1}{2}}^{\sigma}(3) & = \frac{1}{\sqrt{3}}(\alpha \beta \beta+\beta \alpha \beta+\beta \beta \alpha) \nonumber \\
		\chi_{\frac{1}{2}, \frac{1}{2}}^{\sigma1}(3) & = \sqrt{\frac{1}{6}}(2 \alpha \alpha \beta-\alpha \beta \alpha-\beta \alpha \alpha)  \\  \label{s q3}
		\chi_{\frac{1}{2}, \frac{1}{2}}^{\sigma2}(3) & = \sqrt{\frac{1}{2}}(\alpha \beta \alpha-\beta \alpha \alpha) \nonumber \\
		\chi_{\frac{1}{2},-\frac{1}{2}}^{\sigma1}(3) & = \sqrt{\frac{1}{6}}(\alpha \beta \beta+\beta \alpha \beta-2 \beta \beta \alpha) \nonumber \\
		\chi_{\frac{1}{2},-\frac{1}{2}}^{\sigma2}(3) & = \sqrt{\frac{1}{2}}(\alpha \beta \beta-\beta \alpha \beta) \nonumber \\
		\chi_{1,1}^{\sigma}(2) & = \alpha \alpha \nonumber \\
		\chi_{1,0}^{\sigma}(2) & = \frac{1}{\sqrt{2}}(\alpha \beta+\beta \alpha)  \\   \label{s q2}
		\chi_{1,-1}^{\sigma}(2) & = \beta \beta  \nonumber \\
		\chi_{0,0}^{\sigma}(2) & = \frac{1}{\sqrt{2}}(\alpha \beta-\beta \alpha) \nonumber
	\end{align}
For pentaquark system, the total spin quantum number can be 1/2, 3/2 or 5/2.
Considering that the Hamiltonian does not contain an interaction which can distinguish the third component of the spin quantum number, so the wave function of each spin quantum number can be written as follows
	\begin{align}
		\chi_{\frac{1}{2}, \frac{1}{2}}^{\sigma 1}(5) = &\chi_{\frac{1}{2}, \frac{1}{2}}^{\sigma}(3) \chi_{0,0}^{\sigma}(2) \nonumber\\
		\chi_{\frac{1}{2}, \frac{1}{2}}^{\sigma 2}(5) = &-\sqrt{\frac{2}{3}} \chi_{\frac{1}{2},-\frac{1}{2}}^{\sigma}(3) \chi_{1,1}^{\sigma}(2)+\sqrt{\frac{1}{3}} \chi_{\frac{1}{2}, \frac{1}{2}}^{\sigma}(3) \chi_{1,0}^{\sigma}(2)   \nonumber\\
		\chi_{\frac{1}{2}, \frac{1}{2}}^{\sigma 3}(5) = & \sqrt{\frac{1}{6}} \chi_{\frac{3}{2},-\frac{1}{2}}^{\sigma}(3) \chi_{1,1}^{\sigma}(2)-\sqrt{\frac{1}{3}} \chi_{\frac{3}{2}, \frac{1}{2}}^{\sigma}(3) \chi_{1,0}^{\sigma}(2)\nonumber \\ \nonumber &+\sqrt{\frac{1}{2}} \chi_{\frac{3}{2}, \frac{3}{2}}^{\sigma}(3) \chi_{1,-1}^{\sigma}(2)  \nonumber \\
		\chi_{\frac{3}{2}, \frac{3}{2}}^{\sigma 4}(5) = & \chi_{\frac{1}{2}, \frac{1}{2}}^{\sigma}(3) \chi_{1,1}^{\sigma}(2)  \\
		\chi_{\frac{3}{2}, \frac{3}{2}}^{\sigma 5}(5) = & \chi_{\frac{3}{2}, \frac{3}{2}}^{\sigma}(3) \chi_{0,0}^{\sigma}(2) \nonumber \\
		\chi_{\frac{3}{2}, \frac{3}{2}}^{\sigma 6}(5) = & \sqrt{\frac{3}{5}} \chi_{\frac{3}{2}, \frac{3}{2}}^{\sigma}(3) \chi_{1,0}^{\sigma}(2)-\sqrt{\frac{2}{5}} \chi_{\frac{3}{2}, \frac{1}{2}}^{\sigma}(3) \chi_{1,1}^{\sigma}(2)\nonumber \\
		\chi_{\frac{5}{2}, \frac{5}{2}}^{\sigma 7}(5) = & \chi_{\frac{3}{2}, \frac{3}{2}}^{\sigma}(3) \chi_{1,1}^{\sigma}(2) \nonumber
	\end{align}
		
Similar to constructing spin wave functions, we first write down the flavor wave functions of the $q^{3}$ clusters, which are
\begin{align}
	\chi_{\frac{3}{2}, \frac{3}{2}}^{f}(3) & = u u u \nonumber \\
	\chi_{\frac{3}{2}, \frac{1}{2}}^{f}(3) & = \sqrt{\frac{1}{3}}(u u d+u d u+d u u) \nonumber \\
	\chi_{1,1}^{f 1}(3) & = \frac{1}{\sqrt{6}}(2 u u c -u c u- c u u) \nonumber \\
	\chi_{1,1}^{f 2}(3) & = \frac{1}{\sqrt{2}}(ucu -cuu) \nonumber \\
	\chi_{1,1}^{f 3}(3) & = \frac{1}{\sqrt{3}}(uuc+ucu+cuu) \nonumber \\
	\chi_{1,0}^{f 1}(3) & = \frac{1}{\sqrt{12}}(2 u d c+2 d u c-c d u-u c d-c u d-d c u) \nonumber \\
    \chi_{1,0}^{f 2}(3) & = \frac{1}{\sqrt{4}}(-c d u+u c d-c u d+d c u) \nonumber \\
    \chi_{1,0}^{f 3}(3) & = \frac{1}{\sqrt{6}}(udc+duc+ucd+cud+dcu+cdu) \nonumber \\
	\chi_{1,-1}^{f 1}(3) & = \frac{1}{\sqrt{6}}(2 d d c -d c d- c d d)  \\
    \chi_{1,-1}^{f 2}(3) & = \frac{1}{\sqrt{2}}(dcd -cdd) \nonumber \\
    \chi_{1,-1}^{f 3}(3) & = \frac{1}{\sqrt{3}}(ddc+dcd+cdd) \nonumber \\
	\chi_{\frac{1}{2}, \frac{1}{2}}^{f 1}(3) & = \sqrt{\frac{1}{6}}(2 u u d-u d u-d u u) \nonumber \\
	\chi_{\frac{1}{2}, \frac{1}{2}}^{f 2}(3) & = \sqrt{\frac{1}{2}}(u d u-d u u) \nonumber \\
	\chi_{\frac{1}{2},-\frac{1}{2}}^{f 1}(3) & = \sqrt{\frac{1}{6}}(u d d+d u d-2 d d u) \nonumber \\
	\chi_{\frac{1}{2},-\frac{1}{2}}^{f 2}(3) & = \sqrt{\frac{1}{2}}(u d d-d u d)    \nonumber
\end{align}
Then, the flavor wave functions of $\bar{q}q$ clusters are
\begin{align}
	\chi_{1,1}^{f}(2) & = \bar{d} u \nonumber \\
	\chi_{1,0}^{f}(2) & = \sqrt{\frac{1}{2}}(\bar{d} d-\bar{u} u) \nonumber \\
	\chi_{1,-1}^{f}(2) & = -\bar{u} d  \\
	\chi_{\frac{1}{2}, \frac{1}{2}}^{f}(2) & = \bar{d} c \nonumber \\
    \chi_{\frac{1}{2},-\frac{1}{2}}^{f}(2) & = -\bar{u} c \nonumber
\end{align}
As for the flavor degree of freedom, the isospin of pentaquark systems we investigated in this work is  $I=0$.
The flavor wave functions of pentaquark systems can be expressed as
\begin{align}
	\chi_{0,0}^{f 1}(5) & = \chi_{0,0}^{f}(3) \chi_{0,0}^{f}(2) \nonumber \\
    \chi_{0,0}^{f 2}(5) & = \sqrt{\frac{1}{2}} \chi_{\frac{1}{2}, \frac{1}{2}}^{f }(3) \chi_{\frac{1}{2},-\frac{1}{2}}^{f}(2)-\sqrt{\frac{1}{2}} \chi_{\frac{1}{2},-\frac{1}{2}}^{f }(3) \chi_{\frac{1}{2}, \frac{1}{2}}^{f}(2)  \nonumber \\
	\chi_{0,0}^{f 3}(5) & = \sqrt{\frac{1}{3}} \chi_{1,1}^{f}(3) \chi_{1,-1}^{f}(2)-\sqrt{\frac{1}{3}} \chi_{1,0}^{f}(3) \chi_{1,0}^{f}(2) \nonumber  \\
	&~~+\sqrt{\frac{1}{3}} \chi_{1,-1}^{f}(3) \chi_{1,1}^{f}(2)
\end{align}

For the color-singlet channel (two clusters are color-singlet), the color wave function can be obtained by $1 \otimes 1$
\begin{align}
	\chi^{c} =& \frac{1}{\sqrt{6}}(r g b-r b g+g b r-g r b+b r g-b g r)\nonumber \\
	&\times\frac{1}{\sqrt{3}}(\bar{r}r+\bar{g}g+\bar{b}b)
\end{align}

Finally, we can acquire the total wave functions by combining the wave functions of the orbital, spin, flavor and color parts together according to the quantum numbers of the pentaquark systems.

\subsection{Real-scaling method}
To provide the necessary information for experiments to search for exotic hadron states, the coupling calculation
between the bound channels and open channels is indispensable. As we know that some channels are bound because of the strong attractions of the system. However, these states will decay to the corresponding open channels by coupling to open channels and become resonance states. Besides, some states will become scattering state by the effect of coupling to both the open and closed channels. The stabilization method, which is also called the real-scaling method, is one of the effective ways to look for the genuine resonance states.
This method was proved to be a valuable tool for estimating the energies of long-lived metastable states of electron-atom, electron-molecule, and atom-diatom complexes~\cite{Taylor}. It was firstly applied to quark model by Emiko Hiyama~\cite{Hiyama:2018ukv} to search for $P_c$ states.

\begin{figure}[htbp]
	\centering
	\includegraphics[width=8cm]{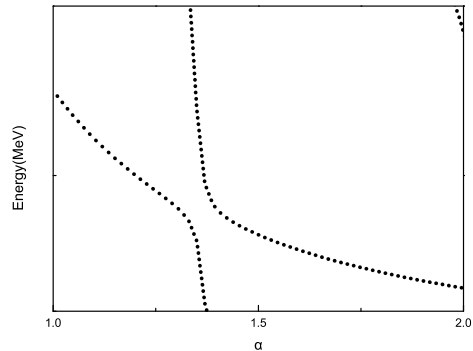}\
	\caption{\label{Figure0} The shape of the resonance in real-scaling method.}
\end{figure}

In this method, a genuine resonance state will act as an avoid-crossing structure (see Fig.~\ref{Figure0}) with the increase of the distance between two clusters, while the continuum states will fall off towards its threshold. In addition, the energy of the bound state will remain unchanged and its figure will behave as a straight line with stable energy.
If the avoid crossing structure is repeated periodically, then the avoid-crossing structure is possible to be a genuine resonance state.
In this case, one can obtain the resonance energy by the corresponding resonance structure.
More details can be found in Refs.~\cite{Simons,Hiyama:2018ukv,Meng:2019fan}.

\section{The results and discussions}
In this work, we investigate the $S-$wave $qqq\bar{q}c$ pentaquark systems in the framework of QDCSM.
The quantum numbers of these systems are $I=0$, $J^P = 1/2^-, 3/2^-$ and $5/2^-$. Two structures $qqq-\bar{q}c$ and $qqc-\bar{q}q$, as well as the coupling of these two structures are taken into account. To find out if there exists any bound state, we carry out a dynamic bound-state calculation. The real-scaling method is employed to investigate the genuine resonance states. Moreover, the calculation of root mean square (RMS) is helped to explore the structure of the bound states or resonance states on the one hand, and to further estimate whether the observed states are resonance states or scattering states on the other hand.

The numerical results of different systems are listed in Tables~\ref{E 0 1/2}, \ref{E 0 3/2} and \ref{E 0 5/2}, respectively.
The first column headed with $Structure$ incluedes  $qqq-\bar{q}c$ and $qqc-\bar{q}q$ two kinds.
The second to forth columns headed with $\chi^{f_i}$, $\chi^{\sigma_{j}}$ and $\chi^{c_k}$ denote the way how wave functions constructed, which can be seen in section II B.
The fifth column headed with $channel$ gives the physical channels involved in the present work.
The sixth column headed with $E_{th}(Theo.)$ refers to the theoretical value of non-interacting baryon-meson threshold.
The seventh column headed with $E_{sc}$ shows the energy of the single channel. The values of binding energies ($E_B$= $E_{sc} -E_{th}(Theo.)$) are listed in the eighth column only if $E_B<0$ MeV.
$E'_{sc}$ is the corrected energy after the mass correction, which will be introduced in detail in the following.
$E_{ccs}$ is the energy of the channel coupling of single spatial structure, while $E_{cct}$ take two spatial structures into account in channel coupling.
$E'_{ccs}$ and $E'_{cct}$ are the corrected energies, corresponding to different coupling modes.
In addition, the proportion of each channel in the channel coupling calculation is shown at the bottom of Tables~\ref{E 0 1/2}, \ref{E 0 3/2} and \ref{E 0 5/2}.
With the help of the proportion of each channel, we can further study the influence of the channel coupling.
	
To reduce the theoretical errors, we can shift the mass to $E'_{sc}=M_{1}(Exp.)+M_{2}(Exp.)+E_{B}$, where the experimental
values of a baryon $M_{1}(Exp.)$ and a meson $M_{2}(Exp.)$ are used.
The above formula is used to handle single channel mass correction.
When we deal with the mass correction of the coupled channels, a modified formula will be used.
$E^{\prime}_{cc}=E_{cc}+\sum_{i} p_{i}\left[E_{\text {th }}^{i}(\text { Exp. })-E_{\text {th }}^{i}(\text { Theo. })\right]$, where $p_{i}$ is
the proportion of various physical channels.

	
\subsection{$J^P=\frac{1}{2}^-$ sector}

\begin{table*}[htbp]
	\caption{\label{E 0 1/2}The energies of the $qqq\bar{q}c$ pentaquark system with quantum numbers $J^P=\frac{1}{2}^-$ (unit: MeV).}
	\begin{tabular}{c c c c c c c c c c} \hline\hline
		~structure~&~~~$\chi^{f_i}$~~~ & ~~~$\chi^{\sigma_j}$~~~ & ~~~$\chi^{c_k}$~~~ & Channel & ~~~$E_{th}(Theo.)$~~~ & $E_{sc}/E'_{sc}$ & ~~~~~$E_{B}$~~~~~ &  ~$E_{ccs}/E'_{ccs}$ &~ $E_{cct}/E'_{cct}$~~  \\ \hline
		\multirow{2}{*}{$qqq-\bar{q}c$}&$i=2 $ & $j=1 $ & $k=1 $ & $ND$ & ~~~2778.3~~~ & ~2779.4/2809.1~ & 0 & \multirow{2}{*}{~2776.4/2801.0~}   & \multirow{7}{*}{~2597.6/2574.4~} \\
		&$i=2 $ & $j=2 $ & $k=1 $ & $ND^*$ & 2862.3 & 2864.4/2946.4 & 0 &   ~  & ~ \\  \cline{1-9}
		\multirow{4}{*}{$qqc-\bar{q}q$}&$i=1 $ & $j=2 $ & $k=1 $ & $\Lambda_{c}\omega$ & 3027.0 & 3029.5/3069.2 & 0 & \multirow{4}{*}{~2613.9/2583.1~} & ~ \\
		&$i=3 $ & $j=1 $ & $k=1 $ & $\Sigma_{c}\pi$ & 2623.0 & 2625.2/2593.6 & 0 &   ~  & ~ \\
		&$i=3 $ & $j=2 $ & $k=1 $ & $\Sigma_{c}\rho$ & 3254.0 & 3251.3/3226.6 & -2.7 &   ~  & ~ \\
		&$i=3 $ & $j=3 $ & $k=1 $ & $\Sigma_{c}^*\rho$ & 3278.6 & 3266.1/3281.2 & -12.5 &   ~  & ~ \\  \hline
		\multicolumn{10}{l}{The proportion of each channel in $qqq-\bar{q}c$ channel coupling, ~$ND:84.0\%$; ~$ND^*:16.0\%$. }   \\
		\multicolumn{10}{l}{The proportion of each channel in $qqc-\bar{q}q$ channel coupling, ~$\Sigma_{c}\pi:87.3\%$; ~$\Lambda_{c}\omega:11.1\%$; ~rests: 1.6\%. }   \\
		\multicolumn{10}{l}{The proportion of each channel in total channel coupling, ~$\Sigma_{c}\pi:90.0\%$; ~$ND:8.1\%$; ~rests: 1.9\%. }   \\ \hline\hline
	\end{tabular}
\end{table*}

The energies of the systems with $J^P=1/2^-$ are listed in Table~\ref{E 0 1/2}, including the $qqq-\bar{q}c$ and $qqc-\bar{q}q$ two spatial structures.
Both single channel and channel coupling results are presented.
First of all, the most intuitive analysis can be based on the results of single-channel calculations.
Obviously, the energies of $ND$, $ND^*$, $\Lambda_{c}\omega$ and $\Sigma_{c}\pi$ channels are above the corresponding theoretical threshold, which means that none of these channels is bound.
The energies of $\Sigma_{c}\rho$ and $\Sigma_{c}^*\rho$ channels are below the corresponding theoretical threshold, with the binding energies -2.7 MeV and -12.5 MeV, respectively.
However, the channels of the system are influenced by each other, so it is unavoidable to take into account the channel coupling effect.

To give a better understanding of channel coupling, we couple the channels with the same spatial structure, as well as all the channels with two spatial structures.
From Table~\ref{E 0 1/2}, we can see that for the $qqq-\bar{q}c$ structure, a bound state with the corrected mass $2801.0$ MeV is obtained, which is close to the charmed baryon $\Sigma_{c}(2800)$. Similar conclusions can be found in Refs.~\cite{Zhao:2016zhf,Dong:2010gu,Zhang:2014ska,Sakai:2020psu,Zhang:2020dwp}. However, this state can decay to the lower channel $\Sigma_{c}\pi$, and it can also be effected by channels with the $qqc-\bar{q}q$ structure.
For the $qqc-\bar{q}q$ structure, the lowest energy of this system is 2613.9 MeV, which is 11.3 MeV lower than the lowest channel $\Sigma_{c}\pi$. So a bound state with the main component $\Sigma_{c}\pi$ is obtained here, and the corrected mass is $2583.1$ MeV. However, it is unclear whether the single bound channel $\Sigma_{c}\rho$ and $\Sigma_{c}^*\rho$ becomes resonance state or not. So the channel coupling of all channels is needed and the real-scaling method is employed to explore the resonance states.

\begin{figure*}[htbp]
	\centering
	\includegraphics[width=18cm]{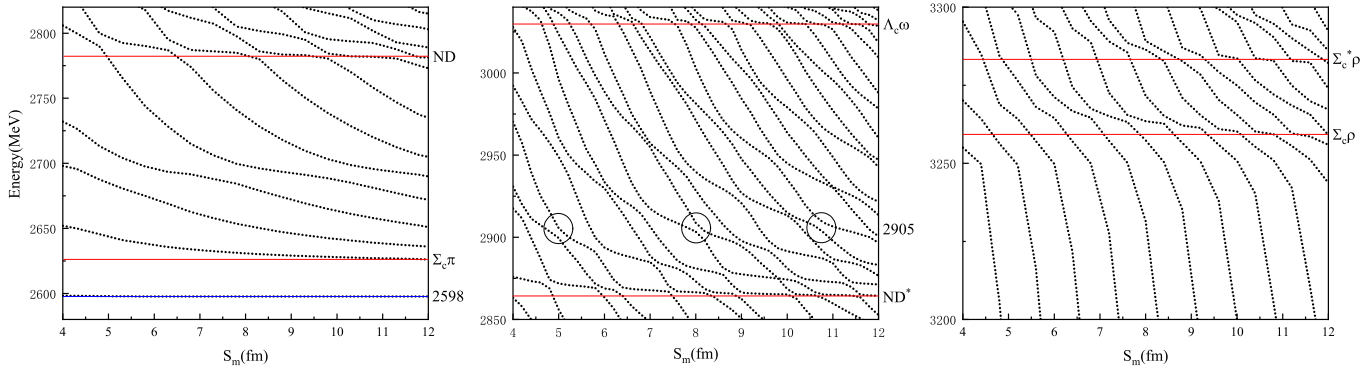}\
	\caption{\label{Figure1} Energy spectrum of $J^P=\frac{1}{2}^-$ system.}
\end{figure*}

In the framework of the real-scaling method, the distance between baryon and meson clusters is labeled as $S_i$, and the largest one is $S_m$. As a result, a genuine resonance state will act as an avoid-crossing structure with the increase of $S_m$, while other continuum states will fall off towards their thresholds. But we have to admit that, when the distance between baryon and meson clusters is small, the threshold structure is not obvious enough. This phenomenon is caused by a lack of computing space and this situation will improve with the increase of $S_m$.
So we calculate the energy eigenvalues of the $qqq\bar{q}c$ system by taking the value of $S_m$ from 4.0 fm to 12.0 fm, to see if there is any stable state. The stabilization plots of the energies of the $qqq\bar{q}c$ system with quantum numbers $J^P=1/2^-$ are shown in Fig.~\ref{Figure1}. The continuum states fall off towards their respective thresholds, which are marked with red lines. And for genuine resonance states, which appear as avoid-crossing structure will be marked with blue lines.
Bound states below the lowest threshold of the systems are also marked with blue lines.

The RMS can also be used to further estimate whether the observed states are resonance states or scattering states. It is worth noting that, the scattering states have no real RMS since the relative motion wave functions of the scattered states are not integrable in the infinite space. If we calculate the RMS of scattering states in a limited space, we can only obtain a value that increases with the expansion of computing space. So we can calculate the RMS of various states to identify the nature of these states by keep expanding the computing space. Besides, the structure of a multi-quark system can also be estimated by calculating the RMS. The results of RMS of the single channel and channel-coupling are listed in the Table~\ref{R 0 1/2}.

After considering the full channel coupling, the lowest energy of the $J^P=1/2^-$ system is pushed down to 2597.6 MeV, and the corrected mass is $2574.4$ MeV, as shown in Table~\ref{E 0 1/2}. After calculating the composition, we find that the channel of $\Sigma_{c}\pi$ is the main component with the proportion of 90.0\%, while the channel of $ND$ and the rest channels are the minor components, accounting for 8.1\% and 1.9\%, respectively. From Fig.~\ref{Figure1} we can see that the energy of this state is very stable with the increase of $S_m$, which confirms that it is a bound state. Besides, the RMS of this state is stable with the increase of the computing space, which further confirms the bound state conclusion. The value of the RMS of this state is 1.35 fm, indicating that the two clusters are not too close to each other. All these properties show that the bound state with $J^P=1/2^-$ is inclined to be a molecular state, and its mass is close the $\Lambda_{c}(2595)$. Since the $\Lambda_{c}(2595)$ is located very close to the $\Sigma_{c}\pi$ threshold, this observation leads us naturally to consider a predominant baryon-meson structure of this lowest-lying odd parity charmed baryon. Here, we prefer to interpreted $\Lambda_{c}(2595)$ as the molecular state with the main component of $\Sigma_{c}\pi$, and the quantum number is $J^P=1/2^-$.
The similar conclusions can be seen in Refs.~\cite{Lu:2014ina,Nieves:2019nol,Zhang:2020dwp}.

As mentioned above, a quasi-bound state is obtained by coupling $ND$ and $ND^*$ channels. From Fig.~\ref{Figure1} we can see that the avoid-crossing structures appear at the place around the threshold of $ND$. However, it is difficult to estimate if there is a resonance state, because the avoid-crossing structure is too close to the threshold of $ND$. We calculate the RMS of the single $ND$ and $ND^*$ channel, as well as the case by channel-coupling (labeled as $E_{ccs}(2776)$). We find that the values of RMS are very large and they are not stable with the increase of the computing space, which indicates that they are scattering states. So we conclude that the avoid-crossing structure around the threshold of $ND$ or $ND^*$ is not a resonance state. It is because the decay rate of different channels to open channels is different, which leads to the different slope of the dots representing energies in the real-scaling figure and forms the avoid-crossing structure. So we cannot explain $\Sigma_{c}(2800)$ as the $ND$ molecular state with $J^P=1/2^-$ in present work.

It is worth noting that in the middle of Fig.~\ref{Figure1}, there exists an avoid-crossing structure around 2905 MeV, the mass of which is close to the newly observed $\Lambda_{c}(2910)$. We're interested in whether this avoid-crossing structure could be a resonance state and explain the $\Lambda_{c}(2910)$.
At the bottom of Table~\ref{R 0 1/2}, we present the proportion of each channel of this $E(2905)$ and its main components are $ND^*$ and $ND$.
Although $E(2905)$ is above the threshold of $ND^*$ and $ND$, it is still possible to be a color-structure resonance state because the effect of the hidden-color channel-coupling is included in the QDCSM. However, the value of RMS of this state is very large and it is variational with the increase of the computing space, which indicates that it is a scattering state. So it cannot be used to explain the $\Lambda_{c}(2910)$ in this work.

For the $\Sigma_{c}\rho$ and $\Sigma_{c}^*\rho$, the single channel calculation shows that both of them are bound states, and the value of RMS of each channel is also consistent with this conclusion. However, after the full channel-coupling, we cannot find any avoid-crossing structure below the threshold of $\Sigma_{c}\rho$ or $\Sigma_{c}^*\rho$. It is reasonable. There are several channels below $\Sigma_{c}\rho$ and $\Sigma_{c}^*\rho$, which will push the energy of these two states above the thresholds. So these two bound states disappear after taking into account the effect of channel-coupling.

\begin{table}[htbp]
	\caption{\label{R 0 1/2}The RMS of the $qqq\bar{q}c$ pentaquark system with quantum numbers $J^P=\frac{1}{2}^-$ (unit: fm).}
	\begin{tabular}{c c c c} \hline\hline
		&~~~~~~~Channel~~~~~~~    & ~~~~~~R~~~~~~ & ~~~~~ nature ~~~~~ \\ \hline
		&$ND$                 &  2.85       &  scattering      \\
		&$ND^*$               &  3.73       &  scattering      \\
		single  &$\Lambda_{c}\omega$  &  3.99       &  scattering      \\
		channel &$\Sigma_{c}\pi$      &  3.39       &  scattering     \\
		&$\Sigma_{c}\rho$     &  1.87       &  bound           \\
		&$\Sigma_{c}^*\rho$   &  1.63       &  bound          \\     \hline
		channel- &$E_{cct}(2598)$       &  1.35       &  bound        \\
		coupling &$E_{ccs}(2776)$      &  3.45       &  scattering    \\
		&$E(2905)$            &  4.33       &  scattering      \\    \hline
		\multicolumn{4}{l}{The proportion of each channel in $E_{cct}(2598)$,  }   \\
		\multicolumn{4}{l}{~$\Sigma_{c}\pi:90.0\%$; ~$ND:8.1\%$; ~rests: 1.9\%. }   \\
		\multicolumn{4}{l}{The proportion of each channel in $E_{ccs}(2776)$,  }   \\
		\multicolumn{4}{l}{~$ND:84.0\%$; ~$ND^*:16.0\%$.}   \\
		\multicolumn{4}{l}{The proportion of each channel in E(2905),  }   \\                                                                       \multicolumn{4}{l}{~$ND^{*}:86.6\%$; ~$ND:10.9\%$; ~rests: 2.5\%. }   \\ \hline\hline
	\end{tabular}
\end{table}

\subsection{$J^P=\frac{3}{2}^-$ sector}
The energies of $qqq\bar{q}c$ pentaquark system with quantum numbers $J^P=\frac{3}{2}^-$ are listed in the Table~\ref{E 0 3/2}.
For the $qqq-\bar{q}c$ spatial structure, the $ND^*$ channel is bound in the single channel calculation.
At the same time, for the $qqc-\bar{q}q$ structure, the single channel calculation shows that both the $\Sigma_{c}\rho$ and $\Sigma_{c}^{*}\rho$ are bound states, with the binding energy of $-76.6$ MeV and $-13.9$ MeV, respectively, while the $\Lambda_{c}\omega$ and $\Sigma_{c}^{*}\pi$ channels are unbound. After coupling all possible channels, a bound state is obtained, whose energy is 2624.4 MeV (2634.7 MeV after mass correction).
The proportion of each channel is $\Sigma_{c}^*\pi:~95.6\%$, $\Lambda_{c}\omega:~2.9\%$ and the rests:~1.5\%, which means that the main component of this bound state is $\Sigma_{c}^*\pi$.

\begin{table*}[htbp]
	\caption{\label{E 0 3/2}The energies of the $qqq\bar{q}c$ pentaquark system with quantum numbers $J^P=\frac{3}{2}^-$ (unit: MeV).}
	\begin{tabular}{c c c c c c c c c c} \hline\hline
		~structure~&~~~$\chi^{f_i}$~~~ & ~~~$\chi^{\sigma_j}$~~~ & ~~~$\chi^{c_k}$~~~ & Channel & ~~~$E_{th}(Theo.)$~~~ & $E_{sc}/E'_{sc}$ & ~~~~~$E_{B}$~~~~~ &  ~$E_{ccs}/E'_{ccs}$ &~ $E_{cct}/E'_{cct}$~~  \\ \hline
		$qqq-\bar{q}c$&$i=2 $ & $j=4 $ & $k=1 $ & $ND^*$ & ~~~2862.3~~~ & ~2853.9/2937.9~ & -8.5 & ~2853.9/2937.9~ &  \multirow{6}{*}{~2624.4/2634.7~}   \\  \cline{1-9}
     	\multirow{5}{*}{$qqc-\bar{q}q$}	&$i=1 $ & $j=4 $ & $k=1 $ & $\Lambda_{c}\omega$ & 3027.0 & 3030.3/3069.2 & 0 & \multirow{5}{*}{~2636.9/2646.9~}    & ~ \\
		&$i=3 $ & $j=4 $ & $k=1 $ & $\Sigma_{c}\rho$ & 3254.0 & 3177.4/3152.7 & -76.6 &~  & ~ \\
		&$i=3 $ & $j=5 $ & $k=1 $ & $\Sigma_{c}^*\pi$ & 2647.6 & 2649.8/2658.0 & 0 & ~  & ~ \\
		&$i=3 $ & $j=6 $ & $k=1 $ & $\Sigma_{c}^*\rho$ & 3278.6 & 3264.7/ 3279.8 & -13.9 & ~  & ~ \\  \hline
		\multicolumn{10}{l}{The proportion of each channel in $qqc-\bar{q}q$ channel coupling, ~$\Sigma_{c}^*\pi:83.8\%$; ~$\Lambda_{c}\omega:8.5\%$; ~rests: 7.7\%. }   \\
		\multicolumn{10}{l}{The proportion of each channel in channel coupling, ~$\Sigma_{c}^*\pi:95.6\%$; ~$\Lambda_{c}\omega:2.9\%$; ~rests: 1.5\%. }   \\ \hline\hline
	\end{tabular}
\end{table*}

\begin{figure*}[htbp]
	\centering
	\includegraphics[width=18cm]{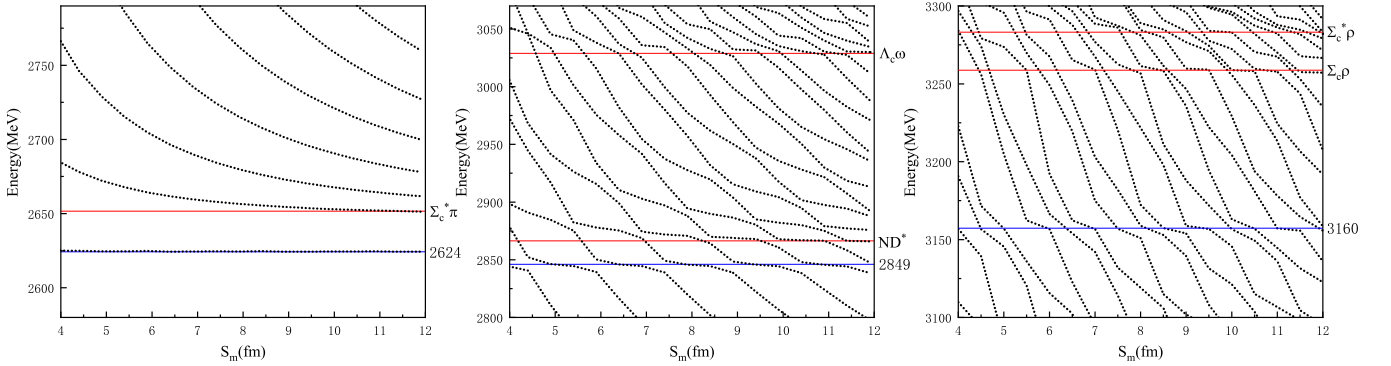}\
	\caption{\label{1.5} Energy spectrum of $J^P=\frac{3}{2}^-$ system.}
\end{figure*}

Fig.~\ref{1.5} and Table~\ref{R 0 3/2} show the stabilization plots of the energies and the RMS of the $qqq\bar{q}c$ system with $J^P=3/2^-$, respectively. It is obvious in Fig.~\ref{1.5} that there is a stable state under the lowest threshold, which is marked by the blue line. The value of RMS of this state is stable with the increase of the computing space and it is 1.44 fm, indicating that the two clusters are not too close to each other. All these properties show that the bound state with $J^P=3/2^-$ tends to be a molecular state, and its mass is close the $\Lambda_{c}(2625)$. So it is possible to interpreted the $\Lambda_{c}(2625)$ as a molecular state with $J^P=3/2^-$ dominated by $\Sigma_{c}^{*}\pi$ channel. The similar explanation could be found in Refs.~\cite{Garcia-Recio:2008rjt,Romanets:2012hm,Zhang:2020dwp}.

In Fig.~\ref{1.5}, there are five red lines, which represent the thresholds for each single channel.
A little below the threshold line of $ND^*$, we can see that stable avoid-crossing structures repeated periodically there. After calculating the composition, we find that the main component is the $ND^{*}$ channel with the proportion of 66.5\%, while the proportion of the $\Sigma_{c}^{*}\pi$ channel is 25.1\% and the one of rest channels is 8.4\%. The theoretical energy of this structure is 2849 MeV, lower than the threshold of $ND^*$. So these avoid-crossing structures may represent a resonance state. By using the proportion of each channel, the corrected mass 2933 MeV is obtained for this state. Besides, the calculation of the RMS of this state shows that it is stable with the increase of the computing space and the RMS is 1.86 fm, which confirms that it is a resonance state with the molecular structure. Clearly, the corrected mass of this resonance state is close to the $\Lambda_{c}(2940)$. So the $\Lambda_{c}(2940)$ is likely to be interpreted as a molecular state with $J^P=3/2^-$, and the main component is $ND^{*}$. This conclusion is consistent with the work of Refs.~\cite{He:2006is,He:2010zq,Dong:2010xv,Dong:2009tg,Ortega:2012cx,Zhang:2012jk,Zhang:2014ska,Wang:2015rda,Entem:2016lzh,Zhao:2016zhf}.

Particularly, there is another repeated avoid-crossing structure below the threshold of $\Sigma_{c}\rho$. The calculated mass of this state is 3160 MeV, and the main component is the $\Sigma_{c}\rho$ channel with the proportion of about 72\%. The corrected mass is 3140 MeV, and the RMS of this state is 1.38 fm. All these properties show that it is also a resonance state, which is worth searching in future work.


\begin{table}[htbp]
	\caption{\label{R 0 3/2}The RMS of the $qqq\bar{q}c$ pentaquark system with quantum numbers $J^P=\frac{3}{2}^-$ (unit: fm).}
	\begin{tabular}{c c c c} \hline\hline
		&~~~~~~~Channel~~~~~~~    & ~~~~~~R~~~~~~ & ~~~~~ nature ~~~~~ \\ \hline
		&$ND^*$                       &  1.91 &  bound       \\
		single &$\Lambda_{c}\omega$    &  3.92   &  scattering      \\
		channel &$\Sigma_{c}\rho$            &  1.40    &  bound     \\
		&$\Sigma_{c}^*\pi$       &  3.56   &  scattering     \\
		&$\Sigma_{c}^*\rho$          &  1.74     &  bound    \\ \hline
		channel-&$E_{cct}(2624)$                  &  1.44    &  bound      \\
		coupling &$E(2849$)                  &  1.86   &  resonance     \\
		   &$E(3160$)             &  1.38   &  resonance     \\  \hline
		\multicolumn{4}{l}{The proportion of each channel in $E_{cct}(2624)$,  }   \\
		\multicolumn{4}{l}{~$\Sigma_{c}^*\pi:95.6\%$; ~$\Lambda_{c}\omega:2.9\%$; ~rests: 1.5\%. }   \\
		\multicolumn{4}{l}{The proportion of each channel in E(2849),  }   \\
		\multicolumn{4}{l}{~$ND^*:66.5\%$; ~$\Sigma_{c}^*\pi:25.1\%$; ~rests: 8.4\%. }   \\
		\multicolumn{4}{l}{The proportion of each channel in E(3160),  }   \\
		\multicolumn{4}{l}{~$\Sigma_{c}\rho:71.9\%$; ~$\Sigma_{c}^*\rho:25.2\%$; ~rests: 2.9\%. }   \\
		 \hline\hline
	\end{tabular}
\end{table}

\subsection{$J^P=\frac{5}{2}^-$ sector}
\begin{table*}[htbp]
	\caption{\label{E 0 5/2}The energies of the $qqq\bar{q}c$ pentaquark system with quantum numbers $J^P=\frac{5}{2}^-$ (unit: MeV).}
	\begin{tabular}{c c c c c c c c c c} \hline\hline
		~structure~&~~~$\chi^{f_i}$~~~ & ~~~$\chi^{\sigma_j}$~~~ & ~~~$\chi^{c_k}$~~~ & Channel & ~~~$E_{th}(Theo.)$~~~ & $E_{sc}/E'_{sc}$ & ~~~~~$E_{B}$~~~~~ &  ~$E_{ccs}/E'_{ccs}$ &~ $E_{cct}/E'_{cct}$~~  \\ \hline
		$qqc-\bar{q}q$ &$i=3 $ & $j=7 $ & $k=1 $ & $\Sigma_{c}^*\rho$ & ~~~3278.6~~~ & ~3173.2/ 3188.3~ & -105.4 & ~3173.2/3188.3~ &  \multirow{1}{*}{~3173.2/3188.3~}   \\ \hline\hline
	\end{tabular}
\end{table*}
For the $qqq\bar{q}c$ system with $J^P=\frac{5}{2}^-$, since only $S$-wave channels are considered in present work, there is only one channel $\Sigma_{c}^*\rho$, which is presented in Table~\ref{E 0 5/2}.
The bound-state calculation shows that it is a deeply bound state, with the binding energy of $-105.4$ MeV. The corrected mass of this state is 3188.3 MeV. The value of RMS of this state is 1.38 fm, which shows that it is also a molecular state.
Although it can decay to some $D$-wave channels, like $ND$, $ND^{*}$, $\Lambda_{c}\omega$, $\Sigma_{c}\rho$, and so on, it is still possible to be a resonance state, which is worthy of experimental search and research.
	
\section{Summary}
In this work, we systematically investigate the $S$-wave pentaquark systems $qqq\bar{q}c$ with $I$ = 0, $J^P$ = $\frac{1}{2}^-,~\frac{3}{2}^- and~\frac{5}{2}^-$ in the quark delocalization color screening model.
The dynamic bound state calculation is carried out to search for any bound state in the $qqq\bar{q}c$ systems. Both the single channel and the channel coupling calculation are performed to explore the effect of the multi-channel coupling.
Meanwhile, the real-scaling method is employed to examine the existence of the resonance states and the bound states.
We also calculate the RMS of cluster spacing to study the structure of the states and estimate if the state is resonance state or not.

The numerical results show that the effect of the channel coupling is important for forming a bound state and deepening the bondage to some extent.
We can draw the following conclusions:
(1) Three bound states are obtained in present work, among which $\Lambda_{c}(2595)$ can be interpreted as the molecular state with $J^P=\frac{1}{2}^-$ and the main component is $\Sigma_{c}\pi$, $\Lambda_{c}(2625)$ can be interpreted as the molecular state with $J^P=\frac{3}{2}^-$ and the main component is $\Sigma_{c}^{*}\pi$. Besides, the $\Sigma_{c}^*\rho$ with $J^P=\frac{5}{2}^-$ is predicted to be a deeply bound state with the mass of 3188.3 MeV. (2) In present work, $\Lambda_{c}(2910)$ cannot be interpreted as a molecular state, and $\Sigma_{c}(2800)$ cannot be explained as the $ND$ molecular state with $J^P=1/2^-$. (3) Two resonance states are obtained, in which the $\Lambda_{c}(2940)$ is likely to be interpreted as a molecular state with $J^P=3/2^-$, and the main component is $ND^{*}$. Besides, a new molecular state $\Sigma_{c}\rho$ with $J^P=3/2^-$ is predicated, whose mass is about 3140 MeV. All these charmed states are worth searching in future work.

In describing the multi-quark system, the channel coupling effect has to be taken into account, especially for the
resonance state, where the coupling to the open channels will shift the mass of the resonance state, or destroy it. 
The real-scaling method may be an effective method to pick up the genuine resonance states from the states with discrete energies. 
Besides, from the above discussion of the charmed baryons, we would
like to note that there exist different points of view to the structure of these states. To explore the
structure of exotic hadrons, the unquenched quark model may be another critical approach.

\acknowledgments{This work is supported partly by the National Science Foundation
of China under Contract Nos. 11675080, 11775118, 11535005 and 11865019 and
Postgraduate Research and Practice Innovation Program of Jiangsu Province under Grant No. KYCX22\_1542.}

\end{document}